\documentstyle[12pt,epsbox]{article}
\begin{document}
\begin{titlepage}
\begin{center}

{\huge   Analysis of immune network dynamical system model
with small number of degrees of freedom}

%{\LARGE --Idiotypic Interaction, Clustering states \\
%and Memory in the Network--}\\
\vspace{1cm}
\large
SSatoko Itaya and Tatsuya Uezu\\
\vspace{.5cm}
 {\em Graduate School of Human Culture,\\
Nara Women's University, Nara 630, Japan\\}

\end{center}
\par

\begin{abstract}
We  numerically study a dynamical system model of an idiotypic
immune network with a small number of degrees
 of freedom.  The model was originally
introduced by Varela et.al, and describes antibodies
 interacting in a body in order to prepare for the invasion
 of external antigens.  \par
  The main purpose of this paper is to
investigate the direction of change in the network system
when antigens invade it.
We investigate three models, the original model, a modified model and
a modified model with a threshold of concentration over which each
antibody can recognize other antibodies.
In all these models, both chaotic and periodic states exist.
In particular, we find peculiar states organized in the
network, the clustering states.

We investigate the response of the system
to invasions by antigens.
We find that in some cases the system
 changes in a positive direction when it is invaded by antigens,
and the clustering state can be interpreted
as memories of the invasion by antigens.
  Further, from the investigation of the relaxation times for
invasions by antigens, it is found  that
in a chaotic state the average response time takes an intermediate
value.
This suggests a positive aspect of chaos in immune networks.
\end{abstract}

\end{titlepage}

\S1. {\bf Introduction}\\
\par
In this paper, we consider
an immune network dynamical system model with small degrees of
freedom.

First, we explain the present understanding of immune
systems briefly\cite{Jerne 74,Varela 91,koyama 96}.\par

The main constituents of an immune system are B-lymphocytes(B-cells)
produced in the Bone marrow, T-lymphocytes(T-Cells) produced in
the Thymus and free antibodies produced by B-cells.
B-cells and T-cells have protein molecules called receptors
on their surfaces.  The receptors of B-cells are
antibodies(Immunoglobulin, Ig), and antibodies recognize
and connect to antigens to neutralize them.
On the other hand, T-cell receptors(TcR) cannot
recognize antigens, but they recognize pieces of antigens
which appear on the surfaces of antigen presenting cells.
When this happens, as a result,
the helper T-cell expedites the immune response
and the suppressor T-cell  suppresses it.  The killer T-cell
attacks and kills a cell which is infected by viruses et.al.
The receptors of B- and T-cells have proper 3-dimensional structures
and these are called 'Idiotypes'.
A family of B-cells which are generated from a B-cell are called
'clones'.  Therefore, a clone and antibodies produced by the clone
have the same idiotype.

In a human body, the total number of clones which are
generated from a single B-cell are about 10 to $10^4$, the total
number of clones amounts to  $10^8$ and the number
of antibodies is about  $10^{20}$.
Thus, the repertoire of antibodies are enough to bind to
any antigen.  This diversity is due to the recombination
and the mutation of genes.

The response to the invasion by antigens is considered as follows.
When antigens enter into a body,  clones which can recognize
the antigen bind to it, and maturate by the help of
the helper T-cells, and a part of them proliferate.
Others become antibody forming cells.
In the antibody forming cells, many antibodies are produced
and secreted.  As a result, many antibodies appear
and neutralize antigens.  When the neutralization completes,
by the action of the suppressor T-cells, the proliferation of B-cells
is suppressed and the immune response ends.\\
Further, in the course of the division of B-cells a part of
each B-cells is preserved as a memory B-cell.
When the same antigens enter into the body again,
these cells rapidly differentiate into antibody forming cells
and produce many antibodies in a short time.
This phenomenon is called 'the secondary immune response'.\par
In the mechanism of the immune response explained above,
a clone of the B-cells which can recognize an antigen
is selected.  Thus, this theory is called 'clonal selection theory'
and has been confirmed experimentally.\par

B-cells, T-cells and antibodies die if they are not stimulated.
As is mentioned above, in a human body there are a huge number
of these cells.  To explain this, in 1974 N. K. Jerne
proposed the so-called network view of the immune systems\cite{Jerne
74}.
In his theory, these cells interact with and activate each other
 organizing a network.\\
However, in 10 years after Jerne's theory
appeared, the network theory was considered
to fail to live up to its initial promise.
This is because the theory cannot explain the correct
direction of change of the system when antigens invade a body.
Another reason is that since
T-cells and their actions were  discovered,
to include these elements the original theory would lose the
simplicity
which attracted many researchers.\par
However, there are several experiments to support
the network theory\cite{exp1,exp2}.  For example, in new born mice
activated lymphocytes are retained although  mice are isolated
from any antigen.  That is,  it seems that the immune system
is activated by itself to prepare for the invasion of external
antigens.
\par

Thus, taking into account T-cells and their roles
F. J. Varela et.al. proposed the second generation immune
network model in 1991\cite{Varela 91}.
Since then, this theory has been developing
\cite{Bersini 94,Varela Stewart 90,Calenbuhr et.al. 94}.\par

We would like to investigate the effects of the interaction
among lymphocytes and antigens, and to analyze what
kind of states and structures the network can have, and to see
which directions the network moves to when antigens
invade the system.
Also, we have interest in  mathematical structures
of  network systems from the view point of dynamical systems.

In this paper, we study the dynamical system model of immune networks
introduced by Varela et.al.
In this model the essential characteristics
of the real immune systems are taken into account.
That is, not only antibodies, but
also  B-lymphocytes which produce antibodies,
and the roles of T-lymphocytes, i.e.,
the activation and the suppression of B-lymphocytes, are included.\par
In reality, an immune system has a huge number
of degrees of freedom. However, in this paper
we focus on the Varela model with small number of degrees of freedom.
Our objective is to investigate the possible states in the network
and the change of these states when antigens invade the system
in small systems, as a necessary step before studying the states
of immune network and the immune response in large systems.
Let us explain the model we treat in this paper.
The constituents of the network, free antibodies and
B-lymphocytes(B-cells), interact with each other through idiotypes.
 Let us distinguish idiotypes by index $i$.
Between two different idiotypes $i$ and $j$, there may occur an
affinity,
 which is represented by the connectivity $m_{ij}$.
We assume  $m_{ij}=1$ if there is an affinity between $i$ and $j$
and $m_{ij}=0$ if not.
$m_{ij}$ is measurable by experiments\cite{exp1,exp2}.
 Let us denote the concentration
of  B-lymphocytes with the $i$-th idiotype by $b_{i}$
 and that of  free antibodies  produced
by the B-lymphocytes by $f_{i}$.
These antibodies have the same idiotype as the
B-lymphocytes.
The sensitivity of the network for the $i$-th idiotype
is defined as follows;
\begin{eqnarray}
\sigma_{i}=\mathop{\sum}_{j=1}^{N}m_{ij}f_{j},
\end{eqnarray}\\
where $N$ is the number of idiotypes.
It represents the strength of the influence by other
antibodies to the $i$-th antibody.
The number of B-lymphocytes and antibodies
change in time by the following causes.
Free antibodies
are removed from the constituents of the network
 because they have a natural lifetime
and also they interact
with other idiotypes and are
neutralized.  On the other hand
they are produced by B-cells as a result of the
 maturation of B-cells.
The probability of the maturation is assumed to depend
on their sensitivity $\sigma$.
This effect is expressed by the function $Mat(\sigma)$.
In the beginning of immune response,
antibodies which interact with antigens
maturate with help of T-cells.  Then, it is natural to assume that
the function $Mat(\sigma)$ is increasing with respect to $\sigma$
when  $\sigma$ is small.
If the number of antibodies becomes large,
and the immune response comes to end, the creation of antibodies
will be suppressed.  Thus, for large values of $\sigma$
$Mat(\sigma)$ should be decreasing with respect to $\sigma$.
Thus, $Mat(\sigma)$ is assumed to have
 the convex profile illustrated
in  Fig.1.

%%%%%%Fig-1%%%%%%%

Then, a differential equation describing
the change in time of the concentration  $f_{i}$
of the $i$-th
antibody can be written as
\begin{eqnarray}
\frac{df_{i}}{dt} &=&
-K_{1}\sigma_{i}f_{i}-K_{2}f_{i}+K_{3}Mat
\left(\sigma_{i}\right)b_{i},
\label{eqn:F}
\end{eqnarray}
where $K_{1}$ is the rate of the neutralization by other antibodies,
$K_{2}$ is the rate of the death of the antibody
 and $K_{3}$ is the rate of
 the creation of the antibodies by B-cells.
Correspondingly B-cells carrying $i$-th idiotype
 on their surfaces decay at a given rate and
proliferate when they maturate.
The probability of the proliferation of B-cells
is represented by the function $Prol(\sigma)$.
When B-cells maturate, they begin to proliferate.
Again, $Prol(\sigma)$ is assumed to be increasing with respect to
$\sigma$ when $\sigma$ is small.
When the neutralization of antigens completes,
the proliferation of B-cells is suppressed
by T-cells.  Therefore, we assume that $Prol(\sigma)$
is decreasing with respect to $\sigma$ when $\sigma$ is large.
Thus, $Prol(\sigma)$ also has the convex shape.
Further, it seems that the proliferation of B-cells ends after their
maturation ends, it is a reasonable assumption that
$Prol(\sigma)$ is shifted to right from $Mat(\sigma)$(Fig.1).\\
Then, the evolution equation for
the concentration $b_{i}$ of the B-cells
with $i$-th idiotype  can be written as
\begin{eqnarray}
\frac{db_{i}}{dt} &=&
-K_{4}b_{i}+K_{5}Prol\left(\sigma_{i}\right)b_{i}+K_{6},
\label{eqn:B}
\end{eqnarray}\\
where $K_{4}$ is the death rate of the B-cells
 and $K_{5}$ is the rate of production
 of the B-cells.  Further, the term $K_{6}$
is added to take into account
the cells that are recruited into the active network
from the bone marrow.

Here, let us see in detail how $Mat(\sigma)$ and $Prol(\sigma)$ work.
First, let us consider $Mat(\sigma)$.

 When the sensitivity $\sigma$ is small, the B-cells are
inactive.  If the value of $\sigma$ becomes large,
 they are activated to maturate by helper T-cells and begin
to produce antibodies.  If the value of $\sigma$
increases further,
the production of antibodies by B-cells
is suppressed by the suppressor T-cells.
As for the behavior of
$Prol(\sigma)$, its behavior is similar to that of
$Mat(\sigma)$.  If sensitivity $\sigma$ becomes large,
the B-cells are proliferated by the T-cells
and then they produce many antibodies.
If $\sigma$ increases further,
this action is suppressed also by the T-cells.

Next, we describe the version of the Varela model introduced
by H. Bersini and B. Calenbuhr,
which we investigate and modify in
this paper.

H. Bersini and V. Calenbuhr\cite{Bersini 94} have investigated
a dynamical system model
of immune networks in the above framework using the following
functions of the maturation and the proliferation
in small degrees of freedom(Fig.2).\\

\begin{eqnarray}
Mat\left(\sigma_{i}\right)&=&\exp\left[ -\left\{\frac{
ln\left(\sigma_{i}/\mu_{m}\right)}{S_{m}}\right\}^{2}\right] \\
Prol\left(\sigma_{i}\right)&=&\exp\left[ -\left\{\frac{
ln\left(\sigma_{i}/\mu_{p}\right)}{S_{p}}\right\}^{2}\right]
\end{eqnarray}\\

%%%%%%%%%Fig-2%%%%%%%%%

In their model, B-cells and antibodies
with the same idiotype are
considered to form a unit.
By the interaction between two units the two-unit system
 oscillates and the phase of
one unit is opposite to that of the other.\\
For a three-unit network, to begin with,
they considered the connection between
two units in an open chain fashion(Fig.3a).
Then, three units can be
constrained with opposite phases each other.

%%%%%%%%%Fig-3%%%%%%%%%

\noindent
Next, they considered the closed network
by connecting three units as shown in Fig.3b.
The connectivity matrix they used is as follows,

\begin{eqnarray}
M=\left[
\begin{array}{ccc}
0 &1 &1\\
1 &0 &1\\
1 &1 &0
\end{array}
\right]
\label{eqn:kihon}
\end{eqnarray}
Although each
pair of units must independently comply
 with the imposed constraint
that they oscillate in opposite phases,
it isn't possible for all units to satisfy this
constraint. This phenomenon has been designated
by the term  ''frustration''  and occurs in a network with
a closed loop composed of an odd number of units.
In general, frustration induces instability.
As a result of this instability,
although the time evolution of each unit resembles
the motion in an open chain network,
this motion does not continue more than several
oscillations, and the network behaves in random way.
In the following section, we investigate
the characteristics of the behaviors in the original model
by Bersini and Calenbuhr.  In \S3,  we modify
the original model by adopting simpler functions
of $Mat(\sigma)$ and $Prol(\sigma)$
and see the effects of the choice of these functions.
In the modified model, we consider a threshold over which
each antibody can recognize others.
This is introduced in \S4.  In \S5, we consider the
networks with more than 3 units and investigate
the effect of degrees of freedom.
Then the invasion of antigens is investigated in \S6.
Finally,  \S7 is devoted to summary and discussions. \\

\S2. {\bf The characteristics of the original model}\\
\par
As is mentioned in the introduction,
the system exhibits periodic oscillations
for both the cases of two units and
of the three-units open chain.
On the other hand, for the case of the three-units
closed chain, the system exhibits
stable chaotic oscillations.

In this system, there always exists the following fixed point

\begin{equation}
 (f_i, b_i)=(0, K_6/K_4),\;\; (i=1, \cdots, N),
\end{equation}
and this point is stable.  However,
this solution has no meaning as a network
because the interaction between units does not exist.

First, to investigate the instability of the system,
we calculated Lyapunov characteristic exponents in the original model
and obtained the following result for the Lyapunov spectrum,
\[
(\lambda_1, \cdots,\lambda_6)=(+,+,0,-,-,-).
\]
This implies that the resultant strange attractor
of this system is rather complicated in structure
because there are two positive exponents.\par
Since this model has permutational symmetry,
it should be checked whether the chaos is robust
with respect to symmetry breaking perturbations.
To investigate this,
we gave random values around 1 to
$m_{ij}$ for $i \ne j$  with $m_{ii}$ fixed to 0,
and studied the time evolution of the system.
We found that chaotic behaviors could also be obtained for
 asymmetric systems.
Thus the chaos in this model is robust
and the cause of appearance of chaos
is not the symmetry of the system.

Next, to see how chaos appears in this system,
we changed
the strength of connectivity retaining
the permutational symmetry in the system.
To be specific, we set the
connectivity matrix as follows,
\begin{eqnarray}
M=s\times\left[
\begin{array}{ccc}
0 &1 &1\\
1 &0 &1\\
1 &1 &0
\end{array}
\right]
\end{eqnarray}\\
and lowered $s$ from 1 to 0.
Then, we obtained the following results. See Fig.4.

%%%%%%%%Fig-4%%%%%%%%%%

Until a rather small value of $s_0\sim 0.4$, the system exhibits
chaotic motion.  When $s$ is less than $s_0$,
a trajectory converges to the fixed point.
%Therefore,  only the chaotic states have
%the significance of the immune network.
Thus,  when the connectivity is small,
this model is not adequate to describe the
immune network.\\
On the other hand, when we increased $s$ from 1,
at $s \sim 1.7$ a limit cycle appears and this breaks the
permutational symmetry of numbering of units.

%%%%%%%%%Fig-5%%%%%%%%%%%

As is shown in Fig.5(a),
the phase portraits of two of the units are the same but
they oscillate in opposite phases to each other.  We call these units
long-pulse units.  One unit has smaller amplitude than those of
the other two units.  We call it the short-pulse unit.
Let us see the characteristic feature of the oscillation of
this limit cycle.  Let us take notice of the time series of antibodies
(Fig.5(b)).
At almost all times, one of the long-pulse units
has a large value of the concentration but
others have small values.
When the largest concentration of a long-pulse unit, say $f_1$,
 becomes small, the other two $f_2$ and $f_3$  increase taking similar
values
initially, then at some value of the concentrations,
the concentration of the other long-pulse unit, say $f_2$,
 becomes large and the short-pulse unit $f_3$ becomes small.
Next, when $f_2$ becomes small,
this time  $f_1$ and $f_3$ increase taking similar value and
 $f_1$ becomes large and $f_3$ becomes small.
Thus, it seems that the short-pulse unit plays a role of 'switching'
of the activation of the long-pulse units.
We call this state a clustering state.\\
As $s$ is increased further, at $s \sim 4.2$.
the permutational symmetry breaks completely and the medium-pulse unit
appears.  This state also can be called a clustering state
and the role of switching is played by the short-pulse unit.\\
At $s \sim 4.7$, the concentrations $f$ and $b$
in one unit are nearly equal to 0.
Since this unit merely affects other units,
we do not regard this state as a clustering state.\par
As for the route to chaos at $s \sim 1.7$,
it is considered to be Intermittency
by heteroclinic intersections
\footnote{It is reported by other authors that when some parameter is
changed, this system shows Intermittency\cite{Calenbuhr et.al. 94}.}.
The sudden disappearance of chaos at $s\sim0.4$
 seems to be crisis, that is, the basin of chaos
intersects with that of stable fixed point.

There are several  characteristic
features of this model by Bersini and Calenbuhr,
 those are,  the topological dimension of the resultant
strange attractor is three,  the chaos is  hyperchaos, and
the switching in the clustering state.
Since in this model the functions $Mat(\sigma)$ and $Prol(\sigma)$
are rather complicated,
it is interesting to investigate
whether these features are due to the special choice of these
functions.
In order to see the effect of these functions,
 we modify the above model by choosing
simpler functions $Mat(\sigma)$ and $Prol(\sigma)$.
In the next section, we go into this study.\\

\S3. {\bf Modified model}\\
\par
We change the functions of the maturation and the proliferation as
follows,
\begin{eqnarray}
Mat\left(\sigma_{i}\right)&=&
U_{1}\times\left[\tanh\left\{U_{2}\times\left(\sigma_{i}-T_{lm}
\right)\right\}\right
.  \\
& & -\left
.\tanh\left\{U_{3}\times\left(\sigma_{i}-T_{um}\right)\right\}
\right] \nonumber , \\
Prol\left(\sigma_{i}\right)&=&U_{4}\times\left[
\tanh\left\{U_{5}\times\left(\sigma_{i}-T_{lp}\right)
\right\}\right .   \\
& & -\left.\tanh\left\{U_{6}\times\left(\sigma_{i}-T_{up}
\right)\right\}\right] \nonumber ,
\end{eqnarray}\\
where $U_1 \sim U_6, T_{lm}, T_{um}, T_{lp}$
and $T_{up}$ are constants.
In Fig.6 we show  the graphs of these functions.  They look rather
similar to those of the previous functions.

%%%%%%%%%%Fig-6%%%%%%%%%%%%%%

When the above functions are adopted
as $Mat(\sigma)$ and $Prol(\sigma)$ ,
in the case of two units, the system converges to
a fixed point and in the case of the three-unit
closed chain, a strange attractor appears.
See Fig.7.

%%%%%%%%%%%Fig-7%%%%%%%%%%%%%%%

The Lyapunov spectrum for chaos becomes as follows.
\[
(\lambda_1, \cdots,\lambda_6)=(+,0,-,-,-,-).
\]
Then, chaos in this system has lower dimension
than that in the original system.\par
To investigate the onset mechanism of chaos,
we drew the bifurcation diagram decreasing
 the magnitude of connection matrix in
the same manner as in the previous section.
See Fig.8.

%%%%%%%%%%%%Fig-8%%%%%%%%%%%%%%

When $s$ is decreased from 1 or increased from 1, the transition from
the strange attractor to a limit cycle takes
place suddenly at $s \sim 0.94$ or $s \sim 1.5$, respectively.
For the limit cycle state which appears below $s \sim 0.94$,
two of three units oscillate in opposite phases,
and the other unit takes negligibly small values.
See Fig.9.

%%%%%%%%%%Fig-9%%%%%%%%%%%%%%%

On the other hand, for the limit cycle state which appears above $s
\sim 1.5$,
two of three units oscillate in opposite phases,
and the other unit oscillates with smaller values.
That is, the permutational symmetry is broken in
these states. The limit cycle state which appears above $s \sim 1.5$
is considered to be a clustering state as discussed in \S2.
See Fig.5 and Fig.10.

%%%%%%%%%%%Fig-10%%%%%%%%%%%%%%%

To see the bifurcation phenomena below $s=1$ in detail,
we calculated the first Lyapunov characteristic
exponent while decreasing the strength $s$ of the
connection matrix.  See Fig.11.\\

%%%%%%%%%%%%%Fig-11%%%%%%%%%%%%%%%%

\noindent
There exists a definite critical point.
In the chaos region, time series of $b_{i}$
exhibit regular oscillations
interrupted by irregular motion.
Thus  the route to chaos
from the limit cycle is Intermittency.
In fact, we confirmed that in both transitions to chaos
taking place at larger and smaller values of $s$ than $s=1$,
 the routes to chaos are
 Intermittency by heteroclinic intersections as in the original
model.\par
As for the robustness,
we checked that the system is robust with respect to
permutational symmetry breaking perturbation
as in the original model.\par
From the results in this section,
we conclude that except for the topological dimensionality of the
chaos,
which is two in this model and three in the original one,
all features are the same as those in the original model.
Thus, the modified model is simpler than the original model.\\

\S4. {\bf The modified model with threshold}\\
\par
In this section, we introduce a threshold over which
antibodies can recognize antibodies and antigens.
 There are several reasons to take
the threshold into account.
One is that it seems that there exists some threshold for
the concentration of antibodies to recognize antigens.
Another reason is that we would like to consider the situation
in which the concentration of an antigen can become large
without being recognized by antibodies for some reason.
As such a situation we can consider the case that
the ability of detecting antigens in the immune system
becomes weak.  Further, as a technical reason,
introducing thresholds makes it possible to define states
clearly.

When the threshold is introduced, the system exhibits various
interesting behaviors. \\

\noindent
\underline{ \bf Threshold}

In the three-unit closed chain system, we
introduce a threshold over which the $i$-th antibody
can recognize other antibodies.
For simplicity we take the common value $f_0$ of thresholds for
all antibodies as follows,
\begin{eqnarray}
\left\{
\begin{array}{ll}
m_{ij}(t)=1  \mbox{  for any $j( \ne i)$}
& \mbox{ when $f_j(t) \ge f_0$}, \nonumber \\
m_{ij}(t)=0  \mbox{  for any $j (\ne i)$}
& \mbox{ when $f_j(t) < f_0$}, \nonumber
\end{array}
\right.
\end{eqnarray}
That is,
\begin{equation}
m_{ij}(t) = m_{ij}\Theta(f_j(t)-f_0),
\end{equation}
where $\Theta(x)$ is the Heaviside function, i.e.,
 $\Theta(x)=1$ for $x \ge 0$ and 0 for $x<0$.
Hereafter, $m_{ij}$ in the right hand side of the above equation
is fixed to the value in eq.(6).
We scanned the value of the threshold $f_0$ every 5 values
and obtained the following behaviors of the system.
\begin{eqnarray}
\begin{array}{ll}
f_{0}=5\sim 15 & \left\{
\begin{array}{l}\mbox{  Limit cycles.
The concentrations of B-cells of two units are}\\
\mbox{ always greater than the threshold and the other is always}\\
\mbox{ less than it.(Fig.12)}
\end{array}
\right. \nonumber \\
f_{0} \sim20 & \mbox{ Limit cycle with period two.}  \nonumber \\
f_{0}=25\sim 40 & \mbox{  Chaos. Fig.13 }  \nonumber \\
f_{0}=45\sim 50 & \mbox{  Limit cycles.
 Clustering state.(Fig.14).}  \nonumber \\
f_{0}\geq 55 & \mbox{  Fixed point.}  \nonumber
\end{array}
\end{eqnarray}
%%%%%%%%%%%Fig-12%%%%%%%%%%%%%%

%%%%%%%%%%%Fig-13%%%%%%%%%%%%%%

%%%%%%%%%%%Fig-14%%%%%%%%%%%%%%

For $f_0=5 \sim 50$, the system is in the clustering state
which is defined in \S2.
In this state, each of three units oscillates taking
values below and over the threshold.  Two long-pulse units
have longer duration over the threshold and the short-pulse unit
has shorter duration.

For the time series analysis, we define the on-off time series
as follows.
Let us associate 0 or 1 with each unit according to the value
of $f_i$, that is, 0 for $f_i < f_0$ and 1 for $f_i \ge f_0$.
We call the former the off-state and the latter
the  on-state,
 and the sequence of 0 and 1 as a function of time
 the on-off time series.
As shown in Fig.15,
at almost all times, there exists only one
on-state, and whenever the long-pulse unit changes
from the on-state
to the off-state, the short-pulse unit becomes the on-state.
That is, the short-pulse unit plays a role of switching.
This is the same phenomenon as observed in \S2 and \S3.

%%%%%%%%%%%%%Fig-15%%%%%%%%%%%%%%

Thus, this phenomenon takes place in all models we studied.
The cause of this phenomenon is ascribed to the nature of
the interaction.
We discuss this in the final section.

If we take an initial state such that
all units are less than the threshold,
the system converges to the fixed point.  On the other hand,
if at least one unit exceeds the threshold initially,
then the system goes to the clustering state.

We investigated the bifurcation structure
when the system is in the clustering state for $f_0=50$.
To see the bifurcation structure clearly,
we introduce the strength of effective interaction $<I_{ij}>$.
$<I_{ij}>$ is defined by the following relation,
\begin{eqnarray}
<I_{ij}>& = & \lim _{T \rightarrow \infty} \frac{1}{T}
\int _0 ^{\infty} dt I_{ij}(t).\\
&& I_{ij}(t) = m_{ij}(t)+m_{ji}(t).
\end{eqnarray}

For example, if $f_1(t) \ge f_0, f_2(t) \ge f_0$
and $f_3(t) < f_0$, then $m_{21}(t)=m_{31}(t)=1,
m_{12}(t)=m_{32}(t)=1$ and $m_{13}(t)=m_{23}(t)=0$.
Thus, $I_{12}(t)=I_{21}(t)=2, I_{13}(t)=I_{31}(t)=1$ and
$I_{23}(t)=I_{32}(t)=1$. See Fig.16.

%%%%%%%%%%Fig-16%%%%%%%%%%%%

As $s$ is lowered, the initial periodic state for $s=1$
becomes chaos at $s\sim 0.93$, and
the permutational symmetry is recovered.
See Fig.17.
  When $s$ is lowered further,
the periodic state appears again at $s\sim 0.52$(Fig.18).\\

%%%%%%%%%%%Fig-17%%%%%%%%%%%%%

%%%%%%%%%%Fig-18%%%%%%%%%%%%%%

\noindent
We show the bifurcation diagram (Fig.19)
and the $s$ dependence of $<I_{ij}>$ (Fig.20).

%%%%%%%%%%%%Fig.19%%%%%%%%%%%%%

%%%%%%%%%%%%Fig-20%%%%%%%%%%%%%

Here, we summarize the result.
\begin{enumerate}
\item  $s_1(\sim 0.93) \le s \le 1$.  Limit cycle. (Clustering state.)
Fig.14.\\
There occurs clustering and there are two groups of $<I_{ij}>$.
That is, the interaction between two long-pulse units,
say $ I_{LL}$,
and the interaction between the long-pulse unit and the short-pulse
unit
, say $ I_{LS}$. Typical values are
\[ I_{LL}\sim 0.83,\;  I_{LS}\sim 0.8. \]
\item  $s_2 (\sim 0.52)\le s \le s_1$.  Chaos. Fig.17.\\
All $<I_{ij}>$ take almost the same values. The typical value is
\[ I_{LL}\sim I_{LS}\sim 0.85. \]
\item  $s_3(\simeq 0.4) \le s \le s_2$.
Limit cycle. Fig.18. \\
As for the usual time series,
one unit takes negligibly small
values and the other two units oscillate in opposite phases.
Then,  the symmetry of this state is broken and two
units always exceed the threshold and the other never
exceeds the threshold.
Reflecting these behaviors, the values of $<I_{ij}>$ are divided into
two groups, - in one group $<I_{ij}>$ tends
to 2 and in the other it tends to 1
as $s \rightarrow s_3$.    Since one unit does not affect other units,
it is not appropriate to call this a clustering state.
\item  $0< s \le s_3$.  Fixed point.\\
There is no meaning of a network.
\end{enumerate}
From these observations, we note that $<I_{ij}>$ can be used
to distinguish the clustering state from other behaviors.
In particular, in the state of chaos, all the $<I_{ij}>$ take
almost the same value.\par
The Lyapunov spectrum for chaos is
\[
(\lambda_1, \cdots,\lambda_6)=(+,0,-,-,-,-),
\]
and is the same as in the modified model.\par
The routes to chaos at  $s\sim 0.93$ and $s\sim 0.52$ are
also Intermittency by the heteroclinic intersections.\\

\S5. {\bf The effect of the degree of freedom}\\
\par
Here we investigate the behaviors of the network
when the number of units is  increased
for the modified model with threshold.
We assume that there exists interactions between
any two units and the
 connectivity matrix is symmetric, i.e., $m_{ij}=m_{ji}=1$
for any $i$ and $j$ $(i\ne j)$.\par
We investigated the cases of  $N=3,4,\cdots,10$.
In Fig.21 and 22,
we show phase portraits for several cases.

The clustering state occurs in all cases.  However, it seems that
it takes place more frequently in the case
of the odd number of units in the parameter ranges which we
investigated.
(Fig.21, 22)

%%%%%%%%%%%%%%%Fig-21%%%%%%%%%%%%%%%%%%

%%%%%%%%%%%%%%%Fig-22%%%%%%%%%%%%%%%%%%

Indeed, for $N=3,5,7$ and 9 and when the threshold is the same
for all units, clustering takes place
and for $N=4, 6, 8$ and 10, it does not.
  In the figures 21(b) and 22(b), we show the effective
interactions schematically.  In these figures
thick lines denote large strength, thin lines represent medium
strength and dotted lines represent small strength.
When the clustering occurs, the system is in a limit cycle state,
and the number of the long-pulse units is larger by one than that of
the
short-pulse units.
There are three values of the strength of the effective interaction in
the system. See Fig.22(b).
The strength of the effective interaction takes the largest value
$I_{LL}$ between two long-pulse units,  the smallest $I_{SS}$
 between two short-pulse units
and intermediate value between the long-pulse unit
and the short-pulse unit.  When the clustering does not occur in the
system,
the strengths of the effective interactions are almost the same.
See Fig.21(b).\\
We show the on-off time series for
$N=3$(Fig.15), $N=5$(Fig.23),
$N=4$(Fig.24(a)) and $N=6$(Fig.24(b)).

%%%%%%%%%%%%%Fig-23%%%%%%%%%%%%%%

%%%%%%%%%%%%%Fig-24%%%%%%%%%%%%

 From  these figures we note that
for $N=3$ and 5, there are only
two types of units, i.e., long-pulse units and short-pulse units.
On the other hand, for $N=4$ and 6, in any unit,
the duration time of the on-state varies in time.
To clarify this we calculate the histogram of the duration time.
See Fig.25, 26, 27, 28.
In the case of odd number of units, we
notice that the duration time $T_L$ for the long pulse unit
is nearly twice as the duration time $T_S$
for the short-pulse unit.  On the other hand,
in the case of even member of units, although
 there are various duration times,
we can find two peaks of the long duration time and the short one.
That is,  in the chaotic state, the role of
each unit changes dynamically.\\

%%%%%%%%%%%%%%%Fig-25%%%%%%%%%%%%%%%%%%

%%%%%%%%%%%%%%%Fig-26%%%%%%%%%%%%%%%%%%

%%%%%%%%%%%%%%%Fig-27%%%%%%%%%%%%%%%%%%

%%%%%%%%%%%%%%%Fig-28%%%%%%%%%%%%%%%%%%

\S6. {\bf The invasion of antigen}\\
Hereafter, we consider the response of the system to the invasion of
antigens in the modified model with threshold.

\par
\noindent
\underline{\bf Case 1}

We consider the clustering state in the  3-units closed network.
Suppose that external antigens
similar to antibodies $f_{1}$ invade the system.
Let us denote the concentration of the antigen by $a_1$ and that
of the corresponding antibodies by $f_1$.
Then, antibodies $f_{2}$ and $f_{3}$ recognize the antigens,
because $a_{1}$ resemble $f_{1}$.
However, in general the antibodies $f_1$ cannot recognize the
antigens.
See Fig.29.

%%%%%%%%%%%%%%%Fig-29%%%%%%%%%%%%%%%%%%

Further, we assume that the antigen does
not proliferate by itself
\footnote{This restricts the type of antigens.
 For example, pollen are one candidate.}.
Thus, the differential equation for the antigen is given by
\begin{eqnarray}
\frac{da_{1}}{dt}=-K_{1}\sigma_a(t) a_{1}+K_{7},
\end{eqnarray}
where $\sigma_a(t)=m_{12}(t)f_{2}(t)+m_{13}(t)f_{3}(t)$.
Here, we assume that the antigens enter into the
system at a rate $K_7$ per unit time.
On the other hand, since antibodies $f_2$ and
$f_3$ interact with the antigens,
$\sigma_2$ and $\sigma_3$
become $\sigma_2 = m_{21}(f_1+a_1)\Theta(f_1+a_1 - f_0)
+m_{23}(t)f_3$
 and $\sigma_3 = m_{31}
(f_1+a_1)\Theta(f_1+a_1 - f_0)+m_{32}(t)f_2$,
respectively.
Then using these $\sigma$s the equations for antibodies and B-cells
are the same as the previous ones. \par
Now, let us see what happens in this system.
The behavior of the system depends on the
increase rate of the antigen $K_7$.
If $K_7$ is large enough, say $K_7>K_7 ^u(\sim 1.2)$,
the concentration of the antigen $a_1$ increases
infinitely and the system is completely invaded
and destroyed by the antigen\footnote{ The system
tends to the fixed point}.
If $K_7$ is less than some value, say  $K_7<K_7 ^l(\sim 0.7)$,
the system copes with the antigens completely.
The number of antigens finally become small and the system
settles near to a clustering state depending on
the initial condition.  Thus, in this case
there is no memory of the invasion by antigens.

If we set  $K_7$ between  $K_7^l$ and $K_7^u$,
e.g., $K_7=0.7$,  $a_1$ does not increase
infinitely but oscillates in some range
of concentration.
The time series of $a_1$ and the phase portrait of the system are
drawn
in Fig.30 and 31, respectively.

%%%%%%%%%%%%%%%Fig-30%%%%%%%%%%%%%%%%%%

%%%%%%%%%%%%%%%Fig-31%%%%%%%%%%%%%%%%%%

\noindent
Although the system is modified because of the
invasion by the antigen, for $K_7^l< K_7< K_7^u$, it still keeps the
nature of the network as a whole.
In the resultant state, the duration time of
the on-state for $f_2$ and $f_3$ are longer,
i.e., the units 2 and 3 are long-pulse units.\par
Now, let us study the response time of the system when
$K_7=0$.
To do this, starting from $a_1=50$ we calculated the relaxation time
in which the concentration of the antigen becomes
negligibly small. See Figure 32.

%%%%%%%%%%%%%%%Fig-32%%%%%%%%%%%%%%%%%%

From this figure, we note that the response time is shorter in
the state in which the units 2 and 3 are
in the long-pulse units.
In the resultant attractor the units 2 and 3 are the long-pulse units.
This result implies
 that the system modifies itself
so as to neutralize the antigen as effectively as possible.
Thus, it seems that the resultant state can respond
much better than other states.  \\
Therefore, we can state
that the resultant attractor is viewed as
a kind of memory of the invasion
by the antigen.\footnote{ We use the term 'memory' to express the
change
to the positive direction of the system under the invasion
by the antigens.  It should not be confused with creation
of memory B-cells.}\par
Further, we scanned the initial value of $A_7$ with $K_7$
fixed to 0.7.  For  $A_1 < 288$, the system settles
to a clustering state.  On the other hand, for $A_1 > 288$
the system tends to a fixed point and the network collapses.
\par
\noindent
\underline{\bf Case 2}

Next, we consider the case that an antigen $A_1$
interacts only with the antibody $f_1$ in the 3-unit closed network.
(Fig.33).\\
%%%%%%%%%%%%%%%Fig-33%%%%%%%%%%%%%%%%%%

\noindent
In this case, we set the thresholds depending on units in order to
obtain clustering states.
Here, we introduce new notations of thresholds, $f_{i,0}( i=1 \sim
3)$,
$g_{1,0}$ and $g_{A,0}$ as follows.
$f_{i,0}$  is the threshold over which the $i$-th antibody
recognizes other antibodies. $g_{1,0}$ is the
 threshold over which the antibody $f_1$ recognizes
the antigen and $g_{A,0}$ is the
 threshold over which the antigen $A_1$
recognizes the antibody $f_1$.
Then the equation for the antigen $A_1$ is
\begin{eqnarray}
\frac{dA_{1}}{dt}&=&-K_{1}\sigma_A(t)A_{1}+K_{7},\\
&&\sigma_A(t)=\Theta(f_1 - g_{1,0})f_1.\nonumber
\end{eqnarray}
The first term expresses the decrease of the antigen
by neutralization by the antibody 1 and the second term represents
the continuous invasion of the antigen.
The sensitivity $\sigma_i$ of the $i$-th unit is modified to
\begin{equation}
\sigma_{i}=\sum_{j=1} ^n m_{ij} \Theta(f_{j}-f_{j,0}) f_j
+ l_i A_1 \Theta(A_1 -g_{A,0}),
\end{equation}
where $l_i$ is the strength of the
interaction between $f_i$ and $A_{1}$.
We put $l_i= s_A \delta _{i,1}$, where $\delta _{i,1}$
is the Kronecker's delta.
The behavior of this network strongly depends
on the thresholds, the value of connectivity $s_A$,
the initial value of the antigen $A_{1}$
and the rate of invasion of the antigen $K_{7}$.
If these values are taken appropriately,
the concentration of the antigen oscillates in some range.
  See Fig.34.\\
%%%%%%%%%%%%%%%Fig-34%%%%%%%%%%%%%%%%%%

%%%%%%%%%%%%%%%%Fig-35%%%%%%%%%%%%%%%%%%

As in the case 1, the unit 1 which can
interact with the antigen is activated and
finally settles near to the long-pulse state.
See Fig.35.

It turns out that the relaxation time of the
concentration of the antigen is comparable in this state
with in the state where
the unit 1 is in the short-pulse state.
See Fig.36.\\

%%%%%%%%%%%%%%%Fig-36%%%%%%%%%%%%%%%%%%

When we compare this result with the result in the case 1,
it is considered that the number of units
which can interact with antigens is important for the
relaxation time. \par

\noindent
\underline{\bf Case 3}

Now, let us consider the network with 4-units.
We assume that the antigen ($ a_1$) has
a similar three-dimensional structure to the
antibody 1 and then can interact
with the unit 2, 3 and 4, but cannot interact
with the unit 1.
See Fig.37.\\

%%%%%%%%%%%%%%%Fig-37%%%%%%%%%%%%%%%%%%

The differential equation for the antigen is
\begin{eqnarray}
\frac{da_{1}}{dt}& = & - K_{1}\sigma_a (t) a_1 + K_{7},\\
&& \sigma_a(t)=(m_{12}(t) f_2 +m_{13}(t)f_3+m_{14}(t)f_4).
\end{eqnarray}
For the unit $i$, the differential equations are
\begin{eqnarray}
\frac{d f_i}{dt}& = & - K_1\sigma_i(t) f_i
-K_2 f_i +K_3 Mat(\sigma_i(t))b_i,\\
\frac{d b_i}{dt}& = & - K_4 b_i
+K_5 Prol(\sigma_i(t)) b_i +K_6,\\
&& \sigma_1(t)=\sum _{j \ne 1} m_{1j}(t)f_j ,\\
&& \sigma_i(t)= m_{i1}(f_1 + a_1) \Theta(f_1 + a_1-f_0)
+ \sum _{j \ne 1} m_{1j}(t)f_j ,\;\; i\ne 1.
\end{eqnarray}

%%%%%%%%%%%%%%%Fig-38%%%%%%%%%%%%%%%%%%

\noindent

If we break the symmetry of the connectivity matrix
by lowering the threshold of one of the four units,
for some initial condition this
unit stays in the short-pulse state
 and other units stay in the long-pulse states,
and so the clustering takes place(Fig.38).
For some other initial conditions, chaotic states appear.
That is, in this case chaos and
the clustering states coexist.

From the result of the case 1, it is expected that
the relaxation time of the system is short
 when the unit which resembles the antigen
is in the short-pulse state.
Further, from the result of the case 2,
it is expected that the more is the number of units
which interact with the antigens, the shorter is the relaxation time.
Thus, here we investigate the relaxation time for the
following three cases.

Case a. The unit 1 is in the short-pulse state.\par

Case b.  The unit 1 is in the long-pulse state.\par

Case c.  The system is chaos.

We show the result in Fig.39.

%%%%%%%%%%%%%%%Fig-39%%%%%%%%%%%%%%%%%%

The relaxation time $\tau _a$ for the case a
is shortest and $\tau _b$ for the case b
is longest.  $\tau _c$ for the case c is
in between  $\tau _a$ and $\tau _b$.
From this result, it is considered that
chaos is more effective than the clustering states
 to prepare for various types of antigens.\\

\S7. {\bf Summary and discussions}\\
\par
\noindent
In this paper, we studied the three models of the immune network
for small number of degrees of freedom $N\sim 10$,
the original model introduced by Bersini et.al.,
the modified model with different functions
of the maturation and the proliferation of B-cells
from those of the original model, and the modified model
with  a threshold over which antibodies can recognize other antibodies
and antigens.  First, we summarize common characteristics
of these models.\par
In these models there exist limit cycle states and chaotic states.
We investigated bifurcation structures obtained by
changing several parameters
and found that the transitions to chaos are Intermittency
by heteroclinic intersections.\par
There is a peculiar type of limit cycle.
In this limit cycle, the permutational symmetry of the system
 is broken and concentrations of antibodies and B-cells
oscillate in a characteristic manner.  We call this the clustering
state.\\
In the clustering state, for example in the
three-unit system,
there are two long-pulse units
in which the concentrations of antibodies are large
and one short-pulse unit in which the concentration
of the antibody is small.
Two long-pulse units oscillate in anti-phase to each other.
At almost all times, only one unit has a
large concentration of the antibody.
  The short-pulse unit
plays a role of ''switching'' the long-pulse units which
take the large concentrations.\par
On the other hand, in the state of chaos,
no such explicit division of roles exist,
but each unit changes its role dynamically.
That is, in chaos, a dynamical
change of roles takes place.\par
We investigated the invasion by antigens
when the clustering takes place.
We found that
when the number of antigens is not too many and interaction between
antigens and antibodies continues for an appropriate period,
the unit which can interact with
antigens settles near to the long-pulse unit after
the concentration of antigens is reduced to small values
in some range of a parameter.\par
By investigating the relaxation time,
we found that in the clustering state
the relaxation time depends on what state
the system stays in.
In the case 1 that the two units interact with antigens
the relaxation time is short or long
when the unit which interacts with antigens is in the long-pulse unit
or short-pulse unit, respectively.
On the other hand, in the case 2 that
only one  unit interacts with antigens
the relaxation time takes similar value both in the long-pulse
unit and in the short-pulse unit.
Thus, if many units can interact with antigens
by the invasion of the antigens
the system moves to the state in which the response to
antigens is most efficient.  Thus, the clustering state is
considered to be a memory of the invasion by antigens.

Further, we checked that the system is robust
against the symmetry breaking perturbations.
Chaotic and periodic states change only a little by these
perturbations.\\

As for the modified model with thresholds, we observed a interesting
feature.
The feature is the positive aspect of chaos
in response to the invasion by antigens.
In this model there is the coexistence of chaos and clustering states
when
the thresholds are chosen appropriately depending on units.
In the response to the antigen invasion,
we found that
the relaxation time in chaotic state
takes an intermediate value between the large
and the small response times in the two types of clustering states.
This suggests a positive aspect of chaos in immune networks,
that is, the possibility that chaos
may cope with the invasion by any kinds of
antigens equally well.\par
Now, let us  discuss the cause of
the appearance of clustering state.
This is related to the mechanism of the switching of long-pulse units
or the dynamical role change.

%%%%%%%%%%%%%%%Fig-40%%%%%%%%%%%%%%%%%%

\noindent
Let us consider the 3-unit closed chain.
See Fig.5(b), 10(b) and 40.
In the present interaction, any two units tend to
oscillate in opposite phases.
If the concentration of the antibody
in one unit, say $f_1$,  increases,
then the sensitivities of  other two units become large.
By this, the second terms of the differential equations for $f_2$ and
$f_3$
become large and $f_2$ and $f_3$ decrease.
This is a kind of 'winner takes all' mechanism.
Then, the sensitivity $\sigma_1$ becomes small.
Thus, when $f_1$ becomes large,  the third term
of the differential equation for $f_1$ becomes small
compared to the other two terms.  Thus,
next $f_1$ begins decreasing.
This causes the decrease of $\sigma_2$ and $\sigma_3$
and the increase of $f_2$ and $f_3$.
Since $f_2$ and $f_3$ begin to increase from small values,
these take similar values.  However, when these become rather large,
two units are forced to stay in states
with opposite phases each other.
Thus, one increases further and the other decreases.
At this stage, for the clustering state,
the increasing unit is fixed to the long-pulse
unit, but for chaos, it is not fixed and the dynamical change of the
switching role takes place.
Therefore, the behavior of the switching is
due to the 'winner takes all' mechanism and the
 'anti-ferro' type interaction between any two units. \par
Next, let us discuss the causes of different behaviors
 between the original model and the modified model without
threshold.\\
We studied influences of each of the terms
in the 3-unit closed chain system.
We pay attention to the unit 1 only,
($f_{1}$,$b_{1}$),
because we consider the case that
the system has the permutational symmetry.
The differential equations of $f_{1}$ and $b_{1}$
are
\begin{eqnarray}
\frac{df_{1}}{dt} &=& -K_{1}\sigma_{1}f_1-K_{2}f_1+K_{3}Mat
\left(\sigma_{1}\right)b_1 \\
\frac{db_{1}}{dt} &=&
-K_{4}b_{1}+K_{5}Prol\left(\sigma_{1}\right)b_1+K_{6}
\end{eqnarray}\\
Here, we denote each term in the above equations as
follows.
\begin{eqnarray}
F_{1} &=& -K_{1}\sigma_{1}f_1,\nonumber \\
F_{2} &=& -K_{2}f_1,\nonumber \\
F_{3} &=& K_{3}Mat\left(\sigma_{1}\right)b_1,\\
B_{1} &=& -K_{4}b_{1},\nonumber \\
B_{2} &=& K_{5}Prol\left(\sigma_{1}\right)b_1,\nonumber \\
B_{3} &=& K_{6}. \nonumber
\end{eqnarray}\\
We compare the time sequence in the original model
with that in the modified model.
See Fig.41,42.
%%%%%%%%%%%%%%%Fig-41%%%%%%%%%%%%%%%%%%

%%%%%%%%%%%%%%%Fig-42%%%%%%%%%%%%%%%%%%

In these figures, solid lines denote
$F_{1}$ and $B_{1}$, large dotted lines represent
$F_{2}$ and $B_{2}$, small dotted lines represent
$F_{3}$ and $B_{3}$, respectively.
We note that the terms concerning the
 maturation and the proliferation give the biggest influence on
 the behaviors of the system in both the models.
In the original model these terms increase and
decrease more rapidly than in the modified model.\\
The chaos in the original model is hyperchaos but that in the modified 
model
is not hyperchaos.
The complexity and the large topological
dimensionality for chaos in the original model
seem to be due to sharp behaviors of these functions.
However, to clarify the influence of the choice of
these functions, it is necessary to perform further investigation.
This is left to a future study.\par
We considered the threshold
over which the antibodies can recognize other antibodies.
We can also consider a threshold over which
the antigen can be recognized by other antibodies.
This is another situation to be investigated.\par
In this study, we investigated
three cases of invasions by antigens.
In these cases, we have interest in
the generic behaviors of the system
when the several parameters are changed, e.g.,
the initial values of antigens or the input
rate of the antigen $K_{7}$, etc.
We found that in some cases the system
changes in a positive way,
that is, in the resultant state the relaxation time becomes
shorter if there are plural units which interact with antibodies.
There exists a kind of memory in the system.
From the  perspective of the real immune network,
the existence of memory states and the response to
the invasion by antigens in a system with
large number of degrees of freedom are very
interesting problems. These will be studied in the future.\\
\par
\begin{center}
{\bf Acknowledgements}\\
\end{center}
The authors are grateful to S. Tasaki, S. Kitsunezaki
and P. Davis for valuable discussions.
One of the authors(S. I.) would like to
thank Professor Y. Kuramoto and the members of his
research group in Kyoto University for valuable discussions.

\newpage
\begin{center}
{\bf Figure captions}\par
\end{center}

Fig.1  Schematic figures for $Mat(\sigma)$ and $Prol(\sigma)$.

Fig.2  $Mat(\sigma)$ and $Prol(\sigma)$ used in the original model.

Fig.3  Type of connection.  open and black circle are in the opposite
phase.\\
\hspace*{2cm}(a) 3-units open chain, (b) 3-units closed chain.

Fig.4 Bifurcation diagram in the original model.\\
\hspace*{2cm}(a) $s \le 1$, (b) $ s \ge 1$.

Fig.5 Periodic solution at $s=1.7 $.\\
\hspace*{2cm}(a) Phase portraits. (b) Time series of $b_i$.

Fig.6  $Mat(\sigma)$ and $Prol(\sigma)$ for the modified model.

Fig.7  Phase portrait of a strange attractor in the modified model.

Fig.8  Bifurcation diagram of the modified model.

Fig.9  Phase portrait of a limit cycle at $s=0.7$.

Fig.10  Limit cycles at $s=1.57$.\\
\hspace*{2cm}(a) Phase portrait.  (b) Time series.

Fig.11 The first Lyapunov exponents.

Fig.12 Phase portrait of a limit cycle at $f_0=10$.

Fig.13 Phase portrait of chaos at $f_0=30$.

Fig.14 Phase portrait of limit cycle at $f_0=50$.

Fig.15 On-off time series in a clustering state.
$s=1, f_0=50$.

Fig.16 Effective interactions.

Fig.17 Chaos at s=0.8.\\
\hspace*{2cm}(a) Phase portrait. (b) On-off time series.

Fig.18 Limit cycle at s=0.4\\
\hspace*{2cm}(a) Phase portrait. (b) On-off time series.

Fig.19 Bifurcation diagram of the modified model with threshold.

Fig.20 $s$ dependence of $<I_{ij}>$.

Fig.21 Attractor in 4-units system in the modified model with
threshold.\\
\hspace*{2cm}(a) Phase portrait. (b) Strength of effective
interactions.

Fig.22 Attractor in 5-units system in the modified model with
threshold.\\
\hspace*{2cm}(a) Phase portrait. (b) Strength of effective
interactions.

Fig.23 On-off time series for $N=5$.

Fig.24 On-off time series. (a) $N=4$. (b) $N=6$

Fig.25 Histogram of duration time. $N=3$.

Fig.26 Histogram of duration time. $N=5$.

Fig.27 Histogram of duration time. $N=4$.

Fig.28 Histogram of duration time. $N=6$.

Fig.29 Schematic figure of interaction between antibodies \\
\hspace*{2cm}and antigen in Case 1.

Fig.30 Time series of antigen $a_1$.

Fig.31 Phase portrait of resultant attractor.

Fig.32 Relaxation times in clustering states.

Fig.33 Schematic figure of interaction between antibodies \\
\hspace*{2cm}and antigen in Case 2.

Fig.34 Time series of antigen $A_1$.

Fig.35 Phase portrait of resultant attractor.

Fig.36 Relaxation times in clustering states.

Fig.37 Schematic figure of interaction between antibodies \\
\hspace*{2cm}and antigen in Case 3.

Fig.38 Clustering state.

Fig.39 Relaxation times in clustering states and in chaos.

Fig.40 Time series of limit cycle at $f_0=50$
in the modified model with threshold.

Fig.41 Time series of each term in the differential equations
of the original model. \\
\hspace*{2cm}(a) $f_1$. (b) $b_1$.

Fig.42 Time series of each term in the differential equations\\
\hspace*{2cm}of the Modified model.(a) $f_1$. (b)  $b_1$.\\

\newpage

\end{document}